\documentclass{IEEEcsmag}
\usepackage[colorlinks,urlcolor=blue,linkcolor=blue,citecolor=blue]{hyperref}
\expandafter\def\expandafter\UrlBreaks\expandafter{\UrlBreaks\do\/\do\*\do\-\do\~\do\'\do\"\do\-}
\usepackage{upmath,color}
\usepackage{graphicx} % Required for inserting images
\usepackage{amsfonts,amssymb,amsmath}            % for math symbols.
\usepackage{adjustbox}
\usepackage[skins,breakable]{tcolorbox}
\usepackage{tikz}
\usepackage{physics}
\usepackage{caption}
\usetikzlibrary{quantikz2}
\tcbuselibrary{listings}
\tcbuselibrary{breakable}

\jvol{XX}
\jnum{XX}
\paper{8}
\jmonth{July/August}
\jname{Publication Name}
\jtitle{Publication Title}
\pubyear{2025}

\begin{document}

\sptitle{Theme Article: Special Issue on Quantum Software and its Engineering}

\title{Quantum Mini-Apps for Engineering Applications: A Case Study}

\author{Horia Mărgărit}
\affil{Quantum Solutions Launchpad, Womanium, Washington DC, USA}

\author{Amanda Bowman}
\affil{Quantum Solutions Launchpad, Womanium, Washington DC, USA}

\author{Geetha Karuppasamy}
\affil{Quantum Solutions Launchpad, Womanium, Washington DC, USA}

\author{Alberto Maldonado-Romo}
\affil{Quantum Solutions Launchpad, Womanium, Washington DC, USA}

\author{Vardaan Sahgal}
\affil{Quantum Solutions Launchpad, Womanium, Washington DC, USA}

\author{Brian J. McDermott}
\affil{Naval Nuclear Laboratory, Niskayuna, NY, 12301, USA}

\markboth{Special Issue on Quantum Software and its Engineering}{Special Issue on Quantum Software and its Engineering}

\begin{abstract}In this work, we present a case study in implementing a variational quantum algorithm for solving the Poisson equation, which is a commonly-encountered partial differential equation in science and engineering. We highlight the practical challenges encountered in mapping the algorithm to physical hardware, and the software engineering considerations needed to achieve realistic results on today's non-fault-tolerant systems.
\end{abstract}

\maketitle

\chapteri{M}odeling and simulation software has been a cornerstone of the design engineer's toolbox since the earliest days of computing. Whether the goal is calculating dynamic loading on a bridge, optimizing the fuel economy of aircraft engine, or maximizing the thermal performance of a CPU heat sink, the primary function of all engineering software is to compute solutions to partial differential equations (PDEs).

This work presents an implementation case study of a variational quantum algorithm (VQA) for solving the Poisson equation in one dimension, a commonly encountered PDE benchmark, and a software architecture for modularizing VQAs to enable ablation studies \cite{Stats_2020}. We review necessary background of the Poisson equation and VQAs; then report our experiences architecting and testing a VQA-based Poisson solver. 
\vspace*{-8pt}

\section{Background}
\subsection{The Poisson Equation}
The Poisson equation (Eq. \ref{eq:poisson_eq}) is an elliptic PDE that describes the equilibrium behavior of a solution field $u$ in the presence of a driving source, $f$. It appears in numerous PDEs of engineering relevance, typically with modifications that introduce time dependence, additional terms, nonlinear couplings, or constitutive relations.
\begin{equation}\label{eq:poisson_eq}
    \nabla^2 u = f
\end{equation}
Consider a heat conduction problem (e.g., CPU and heat sink). $f$ represents the heat source, $u$ represents temperature as a function of position within the system at equilibrium.

Given its ubiquity within computational engineering, the Poisson equation is considered a \emph{"Hello world!"} problem for benchmarking PDE solution methods. These solutions are typically obtained using finite difference or finite element methods.

\subsection{PDE Scaling and Quantum Computing}
All numerical PDE solution methods currently suffer from the \emph{curse of dimensionality} \cite{Bungarts_1999}. As the number of dimensions $d$ increases, current solution methods require $\mathcal{O}(n^d)$ computational resources.

Writing a 3D PDE solver is significantly more challenging than a 2D solver \cite{Ciarlet_2002}, and higher dimensional PDEs like the Boltzmann transport equation or Black-Scholes equations require domain specific, highly tuned solvers.

Quantum computing is emerging as a new architecture because it requires computational resources logarithmic in the total number of grid points. Any PDE solution discretised at $(N=n^d)$ points requires $\mathcal{O}(log N)$ \emph{qubits} for representation on a quantum computer.

Unlike classical bits, \emph{qubits} can be in any complex superposition of states between 0 and 1. A qubit's state is represented by Eq. \ref{eq:qubit}, where the relative contributions of the 0 and 1 states to the complete qubit state $\psi$ are represented by complex coefficients $\alpha, \beta \in \mathbb{C}$ , and $|\alpha|^2 + |\beta|^2 = 1$.
\begin{equation}\label{eq:qubit}
    |\psi\rangle = \alpha|0\rangle + \beta|1\rangle
\end{equation}

The fundamental principles of qubits and quantum computing have been covered elsewhere \cite{Adedoyin_2018}. In this work, we focus on the implications for solving PDEs, specifically how VQAs can significantly reduce the computational overhead compared to classical methods.

The number of iterations for VQAs to converge scales \emph{polylogarithmically} \cite{HHL} with the number of grid points, significantly improving the exponential scaling of classical solvers as the dimensionality increases. Analyzing multi-dimensional problems becomes \emph{tractable} on a quantum computer, suggesting a path to practical quantum advantage in high-dimensional PDE problems.

\subsection{Variational Quantum Algorithms}
While the theory of quantum computing has been extensively developed since its inception \cite{Feynman_1986}, physical implementations of quantum processors have become broadly available to users only in the last decade \cite{IBM}. This computing regime is termed Noisy Intermediate-Scale Quantum (NISQ) computing \cite{Preskill_2018} because these architectures aren't fault tolerant. NISQ hardware susceptibility to errors has driven the development of VQA methods, which are based on the variational principle in quantum mechanics \eqref{eq:variational_method} for finding the ground state $ \ket{\psi^{*}} $ of wavefunctions.

\begin{equation}\label{eq:variational_method}
    E_{0} = \bra{\psi^{*}} H \ket{\psi^{*}} \leq \min_{\ket{\psi}} \frac{\bra{\psi} H \ket{\psi}}{\braket{\psi}}
    = \min_{\ket{\psi}} \bra{\psi} H \ket{\psi}
\end{equation}

Where the constraint in Eq. \eqref{eq:qubit} implies $ \braket{\psi} = 1 $. \newline

Finding the ground state requires minimizing the Rayleigh quotient \eqref{eq:variational_method}, which corresponds to searching for the eigenvalues of matrices that \emph{upper bound} the \emph{lowest energy} $ E_{0} $ of a quantum system whose dynamics are governed by $ H $.

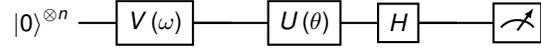
\begin{figure}[h]
    \centering
    \begin{quantikz}
       \\
        \lstick{$\ket{0}^{\otimes n}$} & \gate{V\left( \omega \right)} & 
            \qw & \gate{U\left( \theta \right)} & \gate{H} & \qw & \meter{} \\
    \end{quantikz} 
    \caption{VQA format}
    \label{fig:VQA}
\end{figure}

This entails three steps as demonstrated in Figure Figure \ref{fig:VQA}. Each of these steps is constrained by hardware limitations such as qubit topology and latency between consecutive \emph{shots}, or consecutive measurements of the given quantum circuit.

First, classical data $ \omega $ is encoded into \emph{qubits} using a scheme $ V\left( \omega \right) $ \cite{Rath_2023}. Second, an \emph{ansatz} $ U\left( \theta \right) $ is applied to the encoding scheme, resulting in the \emph{trial wavefunction} $ \ket{\psi} $. Third, a Hermitian operator $ H $ is defined to represent the dynamics of the problem.

VQAs rely on an external, classical optimizer to update the quantum state via the parameterized ansatz. This stands in contrast to quantum imaginary time evolution (QITE) algorithms \cite{Motta_2020}, which iterate through the dynamics of the quantum system to arrive at the ground state. QITE relies on the stability of the Trotter-Suzuki decomposition (TSD), which can lead to convergence issues \cite{Dhand_2014}, and can require more quantum gates than are tractable in the NISQ era. In comparison VQAs are more suitable for NISQ hardware because they bypass the explicit time evolution by directly optimizing ansatz parameters $\theta$, thereby reducing gate depth and making them less vulnerable to hardware noise.

These developments have led to a proliferation in VQA research across many different disciplines of science, engineering, and business. 
\vspace*{-8pt}

\section{Methods}
This section discusses each component of the VQA to solve the Poisson equation \eqref{eq:poisson_eq}, their challenges, and our approach to overcoming them. We additionally discuss the architecture and hardware considerations for implementing the VQA on NISQ hardware. We examined the prior work of Sato et al. \cite{Sato_2021} to solve the Poisson equation using a VQA.

\begin{equation}\label{eq:variational_poisson}
\begin{aligned}
    E_{0} \leq E_{h}\left( r^{*}, \boldsymbol{\theta}^{*} \right) &= \min_{r, \boldsymbol{\theta}} \frac{1}{2} r^{2} \bra{\psi\left( \boldsymbol{\theta}\right)} A \ket{\psi\left( \boldsymbol{\theta}\right)}
    \\
    &- r \bra{f, \psi\left( \boldsymbol{\theta}\right)} X \otimes I^{\otimes n} \ket{f, \psi\left( \boldsymbol{\theta}\right)}
\end{aligned}
\end{equation}

Sato et al. demonstrated that minimizing the Rayleigh quotient \eqref{eq:variational_poisson} provides a symbolic solution to the Poisson equation.

Several challenges arise when implementing this approach on NISQ computing, primarily due to barren plateaus, and the large number of quantum gates required for encoding boundary conditions and for evaluating gradients of the objective function \eqref{eq:variational_poisson}.

We address these challenges by reducing quantum gate counts through gradient-free optimizers, representing boundary conditions via V-Chains or Sparse Pauli Operators, and using tensor network ansatze to mitigate barren plateaus. Additionally, we propose a modular system architecture for the VQA algorithm, enabling fidelity, gate count, and runtime to be evaluated through ablation studies \cite{Vishnusai_2020}.

\subsection{Boundary Conditions}

Quantum computing requires \emph{shift operators} to map PDE boundary conditions to quantum circuits. Implementing shift operators on NISQ hardware is challenging because the gate count they require adds too much noise.

The probability of a correct calculation, $ \rho $, is compounded by an incremental gate's error rate: $ \rho = \rho (1 - \epsilon) $. This exponentially reduces the overall fidelity of the computation. We therefore experimented with V-chains and sparse Pauli operators to reduce gate counts. 

When boundary conditions can be expressed as square matrices, they can be decomposed as a linear combination of Pauli operators:

\begin{equation}\label{eq:shiftop_eq}
    S=\sum_{i,j}C_{ij}P_i \otimes P_j
\end{equation}

where, $P_i$, $P_j$ are Pauli operators, and $C_{ij}$ are the coefficients. This approach reduces the gate count while improving accuracy and stability. However, it requires additional measurements to extract information from quantum states. Multiple measurements require re-initialization of the quantum state and additional communication latency with the external, classical optimizer.

V-chains, unlike Pauli operator decomposition, use ancilla qubits and don't require incremental circuit preparations or measurements. V-chains generally require half of the platform qubits to be reserved as ancilla in order to represent multi-control gates as $ \mathcal{O}\left( 1 \right) $ platform gates. This reduces the size of the Hilbert space through which solutions can be searched to $ \mathcal{O}\left( \sqrt{N} \right) = \mathcal{O}\left( n^{\frac{d}{2}} \right) $ thereby \emph{halving} the PDE dimensions $d$ for which a solution can be found.

\subsection{Avoiding Barren Plateaus}
Consider a neighborhood of coordinates, $ \mathcal{N} $, for an objective function $ f $. A barren plateau occurs when the value of the objective function evaluated at any of the coordinates within $ \mathcal{N} $ is approximately constant: $ f\left( x_{i} \right) \approx f\left( x_{j} \right) \forall i, j : x_{i} \cap x_{j} \in \mathcal{N} $.

The VQA cost function is minimizing the Rayleigh quotient \eqref{eq:variational_method}, which depends on the Hermitian, $ H $, that governs the dynamics of the quantum system, plus the \emph{ansatz} $ V\left( \theta \right) $. Increasing the number of unitary operators in either $ H $ or in $ V\left( \theta \right) $ can exacerbate barren plateaus as these operators weaken the coupling between the local neighborhood $ \mathcal{N} $ and the global landscape, reducing the influence of the global cost function on $ \mathcal{N} $.

To mitigate barren plateaus, we experimented with tensor netwok \emph{ansatze} and with gradient-free optimizers. Our results suggest that tensor networks (MPS, TTN) mitigate barren plateaus while requiring comparable gate counts as hardware efficient ansatze. Our results also suggest that gradient-free optimizers can escape barren plateaus, since calculating quantum gradients is no longer necessary.

Unlike unstructured, randomly initialized ansatze, tensor networks encode entanglement hierarchically and exploit low-rank decompositions to model quantum states with fewer gates. This hierarchy also constrains the expressibility of the ansatz, which has been shown to mitigate the emergence of barren plateaus \cite{Cerezo_Sone_2021}. Matrix Product States (MPS) or Tree Tensor Networks (TTNs) use a linear arrangement of tensors that efficiently capture local correlations in 1D and tree-like structures, respectively, making them suitable for problems with low entanglement entropy. 

Without careful configuration, TTNs can require long-range entanglement between qubits that are physically distant from each other within a quantum system. This requires additional SWAP gates, thereby compounding noise in the measurements.

Fidelity measures the closeness of the quantum state produced by the circuit to the ideal state, and is a standard measure of quantum error. Higher fidelity is desired as it implies minimal measurement error due to quantum noise. Table \ref{Fidelity} reports our experimental results, where MPS ansatze showed consistantly higher fidelity, while TTNs exhibited a notable decrease in fidelity as circuit depth increased. This is attributable to the additional SWAP gates required by TTNs, and we hypothesize TTNs are best suited for all-to-all connected NISQ hardware which wouldn't require additional SWAP gates.

The measurements in Table \ref{Fidelity} were performed on the IBM Osaka quantum computer which realizes 2-local connectivity of qubits. Our results therefore emphasize dependence between hardware architecture and logical circuit representation, and its impact on balancing entanglement representation versus noise resilience.

\begin{table}[h!]
\centering
\begin{tabular}{|>{\centering\arraybackslash}m{3cm}|>{\centering\arraybackslash}m{3cm}|}
\hline
\textbf{Ansatz Type} & \textbf{Fidelity (\%)} \\
\hline
TTN++ & 73 \\
\hline
TTN & 76 \\
\hline
MPS & 80 \\
\hline
custom MPS & 81 \\
\hline
\end{tabular} \\ 
\caption{Fidelity measures of different Ansatz types. The Tree Tensor Network (TTN) ansatz is limited to systems with qubits arranged in powers of 2 (e.g., 2, 4, 8 qubits), whereas extended Tree Tensor Network (TTN++) ansatz extends this architecture enabling it to work with arbitrary qubit numbers. The Matrix Product State (MPS) ansatz allows qubits to be linearly entangled while in the custom Matrix Product State (custom MPS) ansatz, the entanglement pattern is modified by introducing controlled-Z (CZ) gate.}
\label{Fidelity}
\end{table}

Gradient-free optimization techniques such as Nelder-Mead (or Powell's method) can sometimes escape the barren plateau if the area of their simplex (or the line extension in the conjugate direction) is allowed to be large enough such that they sample a point of the cost function outside the barren plateau. These methods require fewer quantum gates and fewer measurements, however, they require an increased number of re-initializations and iterations.

By combining tensor network ansatze with gradient-free optimizers, one can create VQAs that are robust to barren plateaus, thereby making larger quantum circuits practical. 

\subsection{Architecture}
Once a mathematical form of the algorithm has been mapped to a quantum ansatz and Hermitian operator, a significant amount of thoughtful software engineering is required to implement the algorithm in a way that maximizes its flexibility, extensibility, and maintainability.

\begin{figure*}
\centerline{\includegraphics[width=\textwidth]{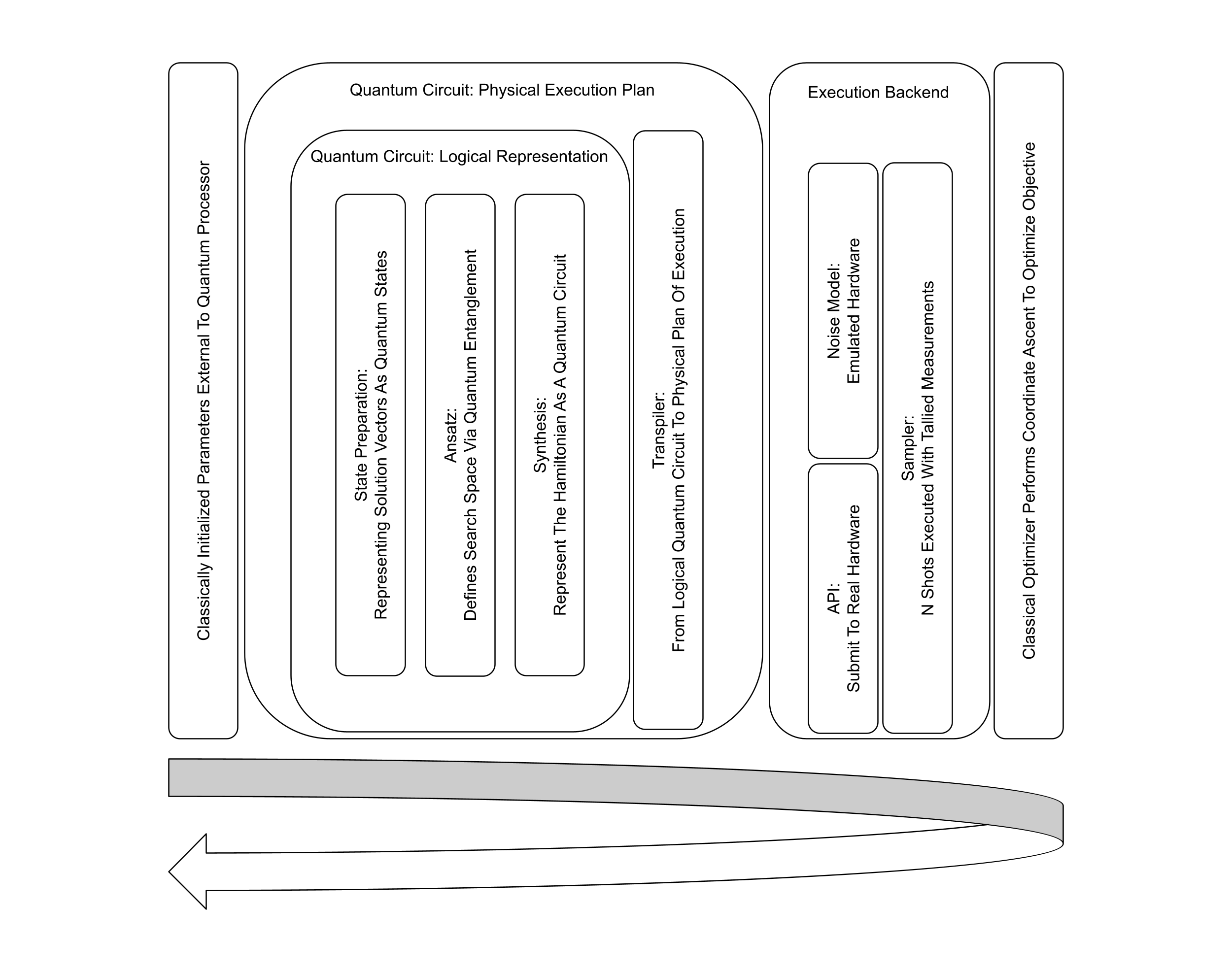}}
\centerline{\includegraphics[width=\textwidth]{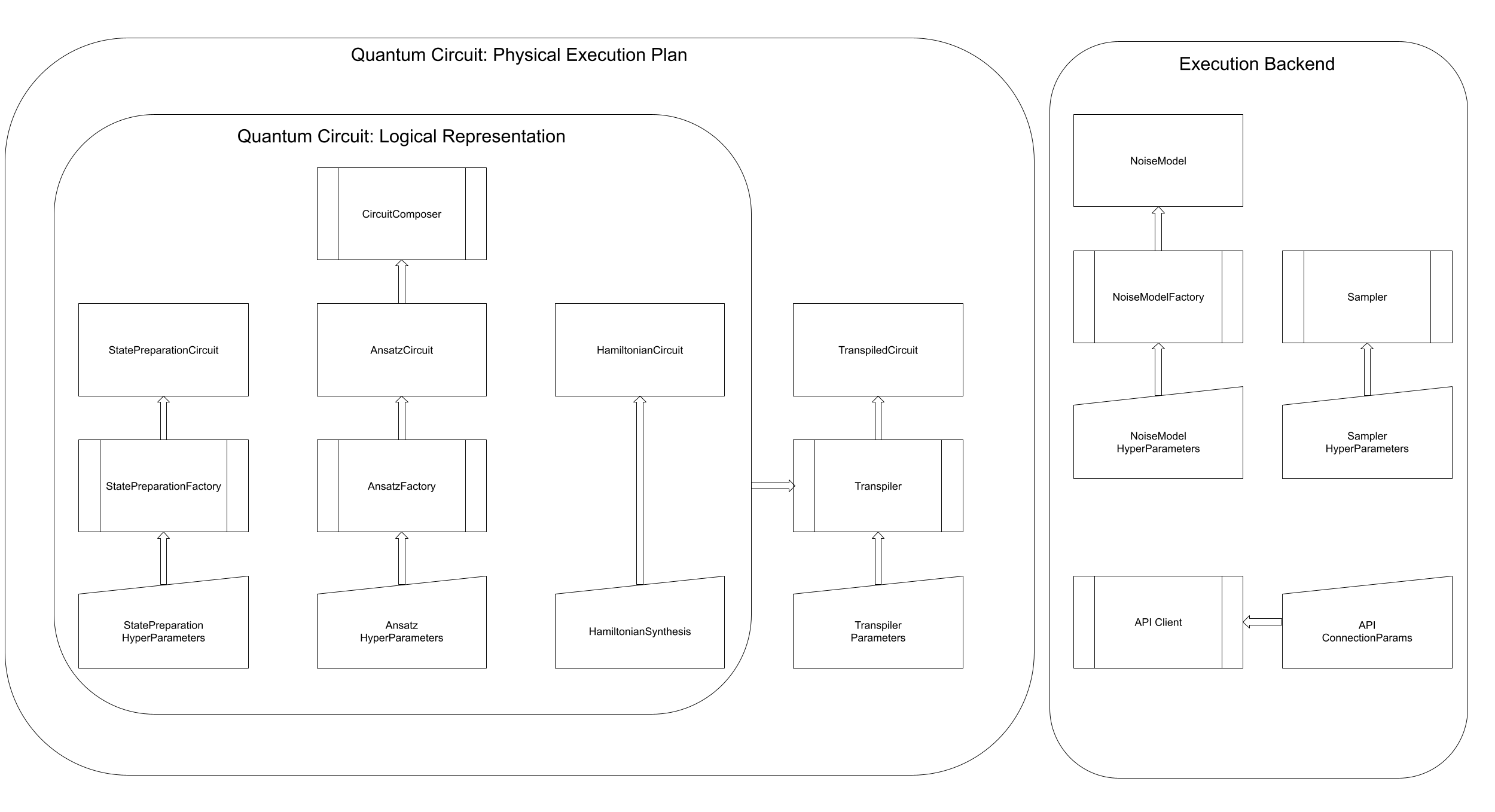}}
\caption{VQA Entity Relationship Diagram}
\label{fig:VQA ERD}
\end{figure*}

The Entity-Relationship Diagram (ERD) (Figure \ref{fig:VQA ERD}) illustrates the core components of our system, and their interactions. This design creates a flexible and modular software stack which supports a clear separation of concerns between quantum circuit design and backend execution.

We decoupled the construction of the quantum circuit from the execution backend. This separation also allows each module to be independently tested and optimized. Additionally, we pipelined the construction of the quantum circuit into two distinct and sequential phases: Logical Circuit Construction, and Physical Circuit Transpilation.

In the first phase, we define a high-level, logical representation of the quantum circuit using standardized building blocks such as gates, qubits, and measurement operations. During this phase, no backend-specific constraints (e.g., hardware connectivity or noise characteristics) are considered, making the logical circuit agnostic to the underlying hardware.

In the second phase, we pass the logical circuit to a transpiler, which maps the high-level representation onto a physical implementation that conforms to the constraints of the selected hardware backend. The transpiler handles tasks such as qubit routing, gate decomposition, and optimization for gate fidelity, generating an executable circuit that is tailored to the specific characteristics of the backend.

Logical circuits can be reused and adapted for different hardware configurations by simply swapping out transpilers or adjusting backend parameters. This layered design enables software practitioners to focus on algorithmic development without being constrained by the hardware-specific details of a particular quantum processor.

We implemented this architecture using the Factory Pattern and Python context managers. There is a factory for each of: noise models, state preparation, and ansatze. Which allows for dynamic instantiation and reuse of common configurations across the software stack.

The sampler, which executes the final circuit, is encapsulated within a context manager. This ensures that the configuration of the sampler, including backend-specific parameters (e.g., shot count, optimization levels), is isolated from the logic defining the quantum circuit. This interface for handling the execution and post-processing of quantum measurements abstracts away additional components like error mitigation routines or advanced measurement strategies from the circuit definition.

The transpiler, which constructs the physical representation of any given logical circuit, is encapsulated within a context manager. This ensures that we can swap out transpilers as needed to execute our circuit on platforms from different vendors.

This modular architecture enables practitioners to experiment with different layers of the software stack — such as changing noise models or ansatz designs — without affecting other components. This facilitates a rapid prototyping workflow, where different quantum algorithms and execution strategies can be tested in isolation before integration.
\vspace*{-8pt}

\section{Results}
\begin{figure}
\centerline{\includegraphics[width=18.5pc]{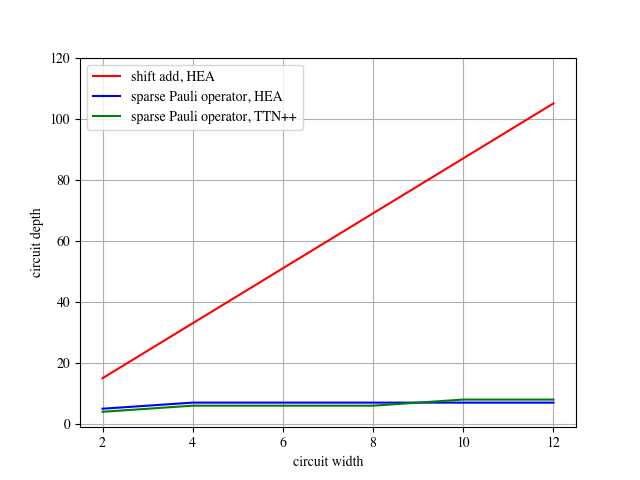}}
\caption{Depth of circuits for the shift add operator used in the Sato et al. implementation (red) and the sparse Pauli operator method using the hardware efficient ansatz (blue) and the TTN++ ansatz (green) with increasing number of qubits for one ansatz layer.}
\label{fig:gate_depth}
\end{figure}

Our experiments focused on optimizing the circuit depth and fidelity of the Sato et al. \cite{Sato_2021} framework by employing various approaches elaborated in the Methods section. While these modifications significantly reduced the overall circuit depth, as depicted in Figure \ref{fig:gate_depth}, the optimizations did not result in a PDE solution competitive to classical solvers. Our study highlights the sensitivity of VQA performance to cost function design, and to poor performing local minima.

\begin{figure}
\centerline{\includegraphics[width=18.5pc]{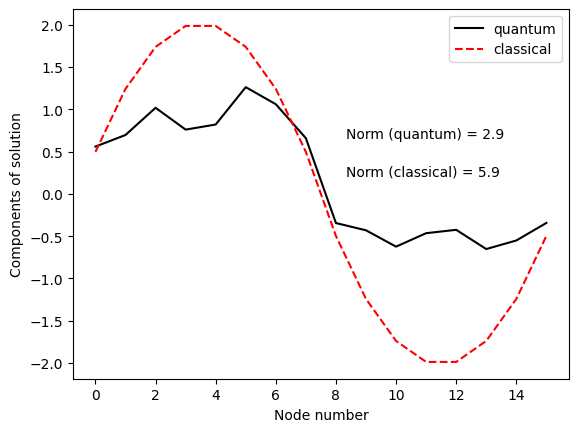}}
\caption{Solution to Poisson equation with periodic boundary conditions using Sato et al. cost function on the QASM simulator with four qubits.}
\label{fig:qasm_result}
\end{figure}

Moreover, our analysis on 4-qubit circuits suggests that the original cost function proposed by Sato et al. in Eq. \ref{eq:variational_poisson} may not be a viable method for ensuring convergence to the correct solution. This is corroborated by the observed instability in the cost landscape. Despite addressing all of the identified limitations in the Sato et al. implementation, including the high circuit depth and inefficient encoding of the boundary conditions, the results obtained on noise-free QASM simulators were still truncated (Figure \ref{fig:qasm_result}), indicating that deep quantum circuits with complex ansatze are prone to convergence issues even in the absence of hardware noise.

Our findings suggest that further exploration into alternative cost functions and hybrid quantum-classical optimization techniques, possibly leveraging shallow, problem-specific ansatze, is necessary to achieve practical quantum advantage for PDE solvers in the NISQ era. 

Additionally, we recommend the structured architecture with separated circuit construction and circuit transpilation modules. Such a system provides several practical benefits for quantum engineers and all software practitioners:

\begin{itemize}
\item[{\ieeeguilsinglright}] {\it Code Reusability:}---The use of factories and context managers allows for easy reuse of components across different projects, reducing boilerplate and simplifying code maintenance.	
\item[{\ieeeguilsinglright}] {\it Backend Agnosticism:}---Decoupling the logical circuit from its physical implementation means that practitioners can focus on algorithmic development without being tied to a specific hardware architecture.
\item[{\ieeeguilsinglright}] {\it Testing and Debugging:}---Each phase (logical construction, transpilation, and execution) can be independently tested, which enhances the reliability and robustness of the software.
\end{itemize}
\vspace*{-8pt}

\section{Conclusion}
In principle, quantum algorithms can help alleviate the curse of dimensionality associated with higher dimensional PDEs. As the fault tolerance of quantum hardware steadily improves in the coming years, it is likely that larger-scale PDE problems of engineering relevance will be demonstrated on these devices. At the present time, implementing quantum algorithms that target hardware backends remains a challenging undertaking, requiring not only an intuition for the quantum mechanical abstractions involved but also a working knowledge of the physical processes occurring on the hardware.
\vspace*{-8pt}

\section{Acknowledgments}
This work was performed as part of the Womanium Quantum Solutions Launchpad program. The authors would like to thank Yury Chernyak for his technical contributions during the initial phase of the project; and Jasmine Murphy, Prachi Vakharia, Marlou Slot, and Shabin Raj for their help and support in managing this project. The team would also like to extend their gratitude towards Jesse Holmes from NNL, and the UMD-IonQ-QLab for their invaluable guidance and assistance throughout the project.

This manuscript has been co-authored by contractors of the U.S. Government under contract number DOE89233018CNR000004. Accordingly, the U.S. Government retains a non-exclusive, royalty-free license to publish or reproduce the published form of this contribution, or allow others to do so, for U.S. Government purposes.
\vspace*{-8pt}
%TC:ignore
\def\refname{REFERENCES}
\vspace*{-8pt}

\begin{IEEEbiography}{Horia Mărgărit}{\,} is a veteran statistician, computer scientist, and a doubly appointed research fellow at the Quantum Solutions Launchpad, Womanium, and the Wehab Lab at Stanford University. His research interests span artificial intelligence and quantum computing for healthcare applications. He has published award winning papers at CHI, serves as a technical committee member of ACHI, and has earned dual bachelors from UC Berkeley and an MS Statistics from Stanford. (contact: \href{mailto:horia@alumni.stanford.edu}{horia@alumni.stanford.edu})
\end{IEEEbiography}

\begin{IEEEbiography}{Amanda Bowman}{\,} is a computational scientist specializing in quantum algorithm research and is a research fellow at the Quantum Solutions Launchpad, Womanium. She holds an MS in computational science from with research in quantum simulations from San Diego State University. 
(contact: \href{mailto:adbowman3@gmail.com}{adbowman3@gmail.com})
\end{IEEEbiography}

\begin{IEEEbiography}{Geetha Karuppasamy}{\,} is a PhD candidate in Computer Science at Oklahoma State University. Her research specialises in Quantum Algorithms and Applications across various computational domains. She is a research fellow at the Quantum Solutions Launchpad, Womanium. (contact: \href{mailto:kkarupp@okstate.edu}{kkarupp@okstate.edu})
\end{IEEEbiography}

\begin{IEEEbiography}{Alberto Maldonado-Romo}{\,} is a PhD candidate in Computer Science at Instituto Politécnico Nacional, Mexico. He leads the Quantum Open Source Foundation (QOSF), is an ambassador for the Unitary Fund and a Qiskit advocate. His research focuses on Quantum Machine Learning, algorithms, error mitigation, software, and education. He is a research fellow at the Quantum Solutions Launchpad, Womanium. (contact: \href{mailto:alberto.maldo1312@gmail.com}{alberto.maldo1312@gmail.com})
\end{IEEEbiography}

\begin{IEEEbiography}{Vardaan Sahgal} {\,} is the Quantum Software \& Solutions head at Quantum Solutions Launchpad, Womanium - where he leads research and development in quantum algorithms and optimization techniques for near-term quantum hardware. His work encompasses designing novel quantum algorithms, improving quantum circuit efficiency, and exploring NISQ-era applications for real world business use-cases. He holds a BS from the University of Delhi and an MS in experimental Material Science from NSUT, Delhi. (contact: \href{mailto:Vardaan.Sahgal@womanium.org}{Vardaan.Sahgal@womanium.org})
\end{IEEEbiography}

\begin{IEEEbiography}{Brian J. McDermott} {\,} is a principal R\&D engineer in the Future Technology group at the Naval Nuclear Laboratory and adjunct professor at Rensselaer Polytechnic Institute.  His work focuses on developing engineering applications of novel and unconventional technologies in computing and applied physics. He holds a Ph.D. in Nuclear Engineering from Rensselaer Polytechnic Institute. (contact: \href{mailto:brianj.mcdermott@unnpp.gov}{brianj.mcdermott@unnpp.gov})
\end{IEEEbiography}
%TC:endignore
\end{document}